\documentclass[twocolumn,showpacs,preprintnumbers,amsmath,amssymb,times]{revtex4}

\usepackage{graphicx}
\usepackage{dcolumn}
\usepackage{bm}

\begin{document}
\title{Urban traffic from the perspective of dual graph}
\author{Mao-Bin Hu$^{1,2}$}\email{humaobin@ustc.edu.cn}
\author{Rui Jiang$^1$}
\author{Yong-Hong Wu$^2$}
\author{Wen-Xu Wang$^3$}
\author{Qing-Song Wu$^1$}\email{qswu@ustc.edu.cn}

\affiliation{$^{1}$School of Engineering Science, University of
Science and Technology of China, Hefei 230026, P.R.China \\
$^2$Department of Mathematics and Statistics, Curtin University of Technology,
Perth WA6845, Australia \\
$^3$Department of Electronic Engineering, City University of Hong Kong,
Hong Kong SAR, P.R. China}

\date{\today}
\begin{abstract}
In this paper, urban traffic is modeled using dual graph representation
of urban transportation network where roads are mapped to nodes and
intersections are mapped to links.
The proposed model considers both the navigation of vehicles on the network
and the motion of vehicles along roads.
The road's capacity and the vehicle-turning ability at intersections
are naturally incorporated in the model.
The overall capacity of the system can be quantified by a phase transition
from free flow to congestion.
Simulation results show that the system's capacity depends greatly
on the topology of transportation networks.
In general, a well-planned grid can hold more vehicles and its overall
capacity is much larger than that of a growing scale-free network.
\end{abstract}
\pacs{45.70.Vn, 89.75.Hc, 05.70.Fh}
\maketitle

\section{Introduction}
Traffic problem is of great importance for the safety and
convenience of modern society
\cite{NS,Helbing,Kerner,BML,CKH,Angel}.
The traffic research was
mainly focused on the highway traffic \cite{NS,Helbing,Kerner} or
the traffic on a well-planned grid lattice \cite{BML,CKH,Angel}.
Recently, empirical evidences have shown that many transportation
systems can be described by complex networks characterized by the
small world \cite{WS} and/or scale-free properties \cite{BA}. The
prototypes include urban road network
\cite{BJiang,Rosvall,Crucitti,Crucitti2,Porta}, nation-wide road
network \cite{Kala}, public transport network \cite{Sien,LiP},
railway network \cite{Sen,LiKP}, airline transport systems
\cite{LiW}, and so on. And due to the importance of large
communication networks such as the Internet and WWW, much
investigation have been focused on ensuring free traffic flow and
avoiding traffic congestion on networks.


In urban traffic, one natural approach is the ``primal"
representation which considers intersections as nodes and road
segments as links between nodes. However, this representation can
not characterize the common notion of ``main roads" in urban traffic
systems.
Moreover, it violates the intuitive notion that an
intersection is where two roads cross, not where four roads begin,
so that it can not represent the way we usually tell each other how
to navigate the road network \cite{Kala}, e.g., ``stay in Manning
Road ignoring 4 crossings, until you reach the Kent Street and turn
right."

Here, we study the urban traffic by employing the dual
representation of the road network, in which a node represents a
single road, and two nodes are linked if their corresponding roads
intersect. This kind of transformation was firstly proposed in the
field of urban planning and design with the name Space Syntax
\cite{Hiller,Hiller2,Penn}, and has been used recently to study the
topological properties of urban roads
\cite{BJiang,Rosvall,Crucitti,Crucitti2,Porta}. In this dual
presentation, the ``main roads" of the city can be represented by
the ``hub nodes" in the network. The node degree is not limited and
it was found that the dual degree distribution of some cities
follows a power-law.

One shortcoming of the dual representation is the abandonment of
metric distance so that a street is one node no matter how long it
is. However, the metric distance is the core of most traffic flow
studies. To avoid this problem, this paper propose a traffic flow
model considering
the movement of vehicles along the road. 
We show that by using two parameters characterizing the road and
intersection capacity, it is possible to simulate the urban traffic
and seek out the overall capacity of the urban transportation
system.

\section{Dual Graph of Urban Network and the Traffic Model}

The basic steps of dual representation can be explained as
following: one should firstly define which road segments belong
together. Previous studies grouped the segments by line of sight by
a driver \cite{Hiller,Hiller2,Penn}, their street name
\cite{BJiang,Rosvall}, or by using a threshold on the angle of
incidence of segments at an intersection
\cite{Crucitti,Crucitti2,Porta}. Then, in the derived ``dual graph"
or the ``connectivity graph", each road is turned into one node,
while each intersection is turned into one link.

\subsection{Characterize Dual Graph Considering Traffic Problem}

With this new paradigm, one can look at urban traffic problem from a
new perspective. From common sense,  the ``main roads" can usually
be characterized by long length, wide road width, many
intersections, and high traffic efficiency. When the main roads are
congested, the entire transportation system will be in danger of
capacity drop. While the system will remain high efficiency even
when some minor roads are jam.

In dual network, the major roads in urban system can be represented
with hub nodes with many links, large capacity, and more handling
ability. In particular, when considering the traffic flow problem,
the characteristics of nodes can be introduced in the dual graph of
urban network:

(1) Degree $k_i$ and degree distribution: The degree (or connectivity) of node $i$
is the number of links connecting to that node.
In the dual graph of urban network, the degree corresponds to the number of
intersections along road $i$.
Thus there is no limit of node degree and so the ``main roads" can be
represented by the ``hub nodes" in the system.

The way the degree is distributed among the nodes can be
investigated by calculating the degree distribution $P(k)$, i.e.,
the probability of finding nodes with $k$ links. Networks with a
power-law distribution are called scale-free \cite{BA}.

(2) Capacity $C_i$: The capacity of road $i$ is the maximum number of
vehicles the road can hold.
This value can be calculated by: $C_i=L_i/l_v$, where $L_i$ is the length
of the road, and $l_v$ is the vehicle's average length.
In the simulation, we assume that the length of each road segment
are same, thus $C_i$ is proportional to the degree of the node:
$C_i=\alpha \times k_i$, where $\alpha$ denotes the maximum number
of vehicles that one road segment can hold.
The system's total capacity is the sum of the capacity of
all node, that is $C_t=\Sigma_i C_i$.

(3) Turning Ability $T_i$: The turning ability is the maximum number
of vehicles turning from the road to neighboring roads per time
step. This value reflects the capacity of intersections along the
road, and can be also measured in real practice. In the dual
representation of urban traffic, $T_i$ is the maximum number of
vehicles a node can send to neighboring nodes. Without losing
generality, we assumed that all intersections can handle the same
number of vehicle-turning from each road, thus $T_i$ is also
proportional to the degree of the node: $T=\beta \times k_i$, where
$\beta$ denotes the ability of one intersection.

\subsection{Traffic Model}

In dual network, the trajectory of a vehicle can be interpreted as
traveling along some roads (nodes) for some distance, and jumping
from a node to another node through an link representing an
intersection. The proposed traffic model considers both the
navigation of vehicles in the network and the movement of vehicles
along roads. The model is partially motivated by the vehicular
traffic models \cite{NS,Helbing,Kerner} and by the traffic models of
packet flow on the Internet \cite{Sole,Arenas,Tadic,Zhao,Wang}. On
the base of underlying dual infrastructure (discussed in Section III
and IV), the system evolves in parallel according to the following
rules:

1. Add Vehicles - At each time step, vehicles are added to the
system with a given rate $R$ at randomly selected nodes and each new
vehicle is given a random destination node.

2. Navigate Vehicles  - If a vehicle's destination is found in its
nearest neighborhood, its direction will be set to the target.
Otherwise, its direction will be set to a neighbor $n$ with probability:
\begin{equation}\label{eq1}
P_n={k^{\phi}_n \over \Sigma_i k^{\phi}_i},
\end{equation}
where the sum runs over all neighbors, and $\phi$ is an adjustable
parameter reflecting the driver's preference with the roads.
It is assumed that the vehicles use a local routing strategy:
they are more ready to use a neighboring ``main road" when $\phi>0$,
and they are more likely to go to a minor road when $\phi<0$.
Once a vehicle reach its destination, it will be removed from the system.

3. Vehicles Motion along Roads -
The intersections along a road $i$ are numbered with serial integers
from $1$ to $k_i$. We use these integers to reflect the sequence of
intersections along the road.
When a vehicle enters a road at the $mth$ intersection and leaves at
the $nth$ intersection, it has to travel the distance of $d = l_0
\times |m-n|$ along the road, where $l_0$ denotes the length of one
road segment, and $|...|$ denotes taking the absolute value. For the
velocity, we assumed that the traffic flow is homogeneous along one
given road. In each time step, the mean velocity of vehicles on the
road is calculated using the following equation:
\begin{equation}
\label{eq2}
v_i=V_{max} \times (1.0-\rho_i),
\end{equation}
where $V_{max}$ is the maximum velocity in the urban system,
and $\rho_i=N_i/L_i$ is the local vehicle density on that road.
Equation (\ref{eq2}) simply approximates that the vehicles will
take $V_{max}$ when there is no vehicle on the road, and the
velocity will decreases linearly until no vehicle can move
when $\rho=1.0$.
In each time step, the distance of each vehicle in the road will
decrease by the amount of $v_i$ until $d=0$, i.e., it reaches the
preferred exiting intersection.

4. Vehicle-Turning at Intersections -
At each step, only at most $T_i$ vehicles can be sent from a node for
the neighboring nodes.
When the number of the vehicles at a selected node is full
(see the introduction of $C_i$), the node won't accept any more
vehicles and the vehicle will wait for the next opportunity.


One can see that the core parameters characterizing the
transportation system are $\alpha$, $\beta$ and $V_{max}$. And
parameter $\phi$ characterizes the behavior of drivers. In the
simulations, without losing generality, the road segment length
$l_0$ is assumed to be $500$ meters, and the average vehicle length
$l_v$ is $10$ meters. Thus we set $\alpha=50$, i.e., one road
segment between two successive intersections can hold $50$ vehicles
at most. The parameter $\beta$ is also set to be constant for all
intersections, meaning that the vehicle-turning ability for each
intersection in the system is the same. And each time step is
assumed to represent $5$ seconds in reality. Thus if the maximum
velocity is $20 m/s$, the maximum velocity in Eq.\ref{eq2} will be
$V_{max}=100$. With changing the values of $\beta$ and $V_{max}$,
different ability of intersections and different road conditions can
be simulated.

\begin{figure}
\scalebox{0.8}[0.8]{\includegraphics{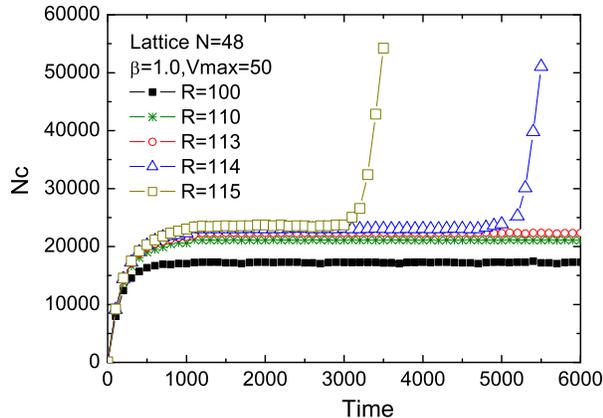}}
\caption{\label{LNc} (Color online)
Typical evolution of vehicle number in the grid system.
The critical generating rate of vehicles is $R_c=113$.}
\end{figure}

\begin{figure}
\scalebox{0.8}[0.8]{\includegraphics{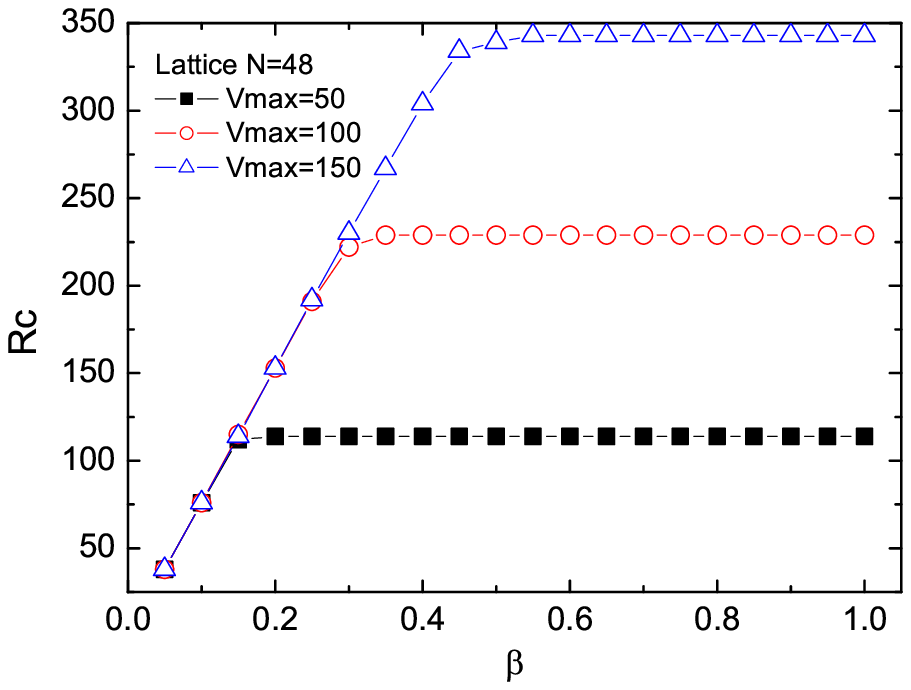}}
\caption{\label{RcB} (Color online)
$R_c$ vs $\beta$ with different values of $V_{max}$.
The data are obtained by averaging $R_c$ with ten simulations.
}
\end{figure}

\begin{figure}
\scalebox{0.8}[0.8]{\includegraphics{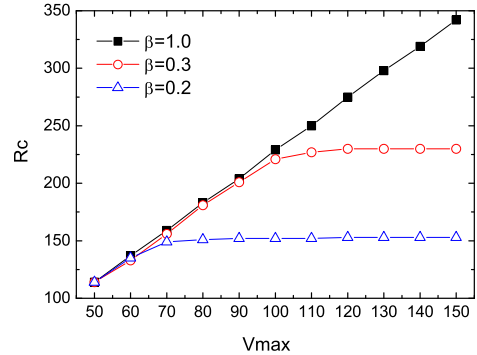}}
\caption{\label{RcV} (Color online)$R_c$ vs $V_{max}$ with
different values of $\beta$ .
The data are obtained by averaging $R_c$ with ten simulations.}
\end{figure}

\section{Simulation of Urban Traffic on a Well-planned Lattice Network}

In this section, the traffic on a well-planned grid urban network is
simulated. The transportation network consists of 24 north-south
roads crossing with 24 east-west roads.  The studied system can be
represented in dual graph by 48 nodes each with the same
connectivity $k=24$. Thus there is no topological ``main roads" in
this system. For the case, $\phi=0$ is applied for all vehicles,
i.e., they randomly select direction from neighboring roads if its
destination is not found. Figure \ref{LNc} displays the typical
evolution of $N_c(t)$, i.e., the number of vehicles within the
system for different generating rate R. One can see that when
$R<R_c$, $N_c$ will increase first and then come to saturation,
indicating the balance of the number of vehicles entering the system
and the number of vehicles reaching their destinations. However,
when $R>R_c$, $N_c$ can not remain constant. It will suddenly
increase and quickly reach the the system's total capacity. Thus the
system is congested and the vehicles accumulate in the system.
Therefore, the critical generating rate $R_c$ can be used to
characterize the phase transition from free flow state to
congestion. And the system's overall traffic capacity can be
measured by $R_c$ above which the system will enter the congested
state.

To better understand the effects of intersection capacity and road
condition on the overall efficiency, simulations with different
values of $\beta$ and $V_{max}$ are carried out. Figure \ref{RcB}
shows the variance of overall capacity $R_c$ with $\beta$. One can
see that $R_c$ firstly increases linearly with $\beta$ and then
comes to saturation when $\beta$ is large enough. And the saturated
value of $R_c$ increases with $V_{max}$. This result is in agreement
with our common sense that we can not improve the overall efficiency
only by enhancing intersection efficiency, but we should also
enhance the road condition so that the vehicles can run faster on
the road.

In Fig.\ref{RcV}, one can see that $R_c$ increases with $V_{max}$
until a saturation is reached. Therefore, when the intersection
capacity is large enough, the road condition will be a limitation
for the whole system; while when the road condition is good enough,
the intersection capacity will be crucial. Unfortunately, the
vehicle speed can not be too large in the city. Thus there will be a
unavoidable limit in improving urban traffic efficiency simply by
enhancing intersection capacity, given that the network topology is
fixed. One should think about other ways, such as adding shortcuts,
developing subways, employing better navigation guidance system for
the drivers, and so on.

\section{Simulation of Urban Traffic on a Self-Organized Scale-Free Network}

Recently, works on the centrality of roads in urban systems using
dual representation
\cite{Rosvall,Sien,Kala,Crucitti,Crucitti2,Porta} show that the
degree distribution of most planned cities is exponential, while it
follows a power-law scaling in self-organized cities. That is, in
most self-organized cities, there are some ``main roads" with many
minor roads intercrossing with them.

In this section, we try to simulate the urban traffic on a dual
graph of scale-free network. To generate the underlying
infrastructure, we adopt the well-known Barab\'{a}si-Albert
scale-free network model \cite{BA}, in which the power-law
distributions of degree is in good accordance with many real
observations. In this model, starting from $m_0$ nodes fully
connected by links with assigned weight $w_0$, the system are driven
by two mechanics: (1) Growth: one node with $m$ links ($m \leq m_0$)
is added to the system at each time step; (2) Preferential
attachment: the probability $\Pi_i$ of being connected to the
existing node $i$ is proportional to the degree $k_i$ of the node
\begin{equation}
\Pi_i={k_i \over \Sigma_j k_j},
\end{equation}
where $j$ runs over all existing nodes.

As a remark, here we do not conclude that the BA network model
exactly describes a self-organized system of urban roads. We adopt
this model to reflect the fact that new roads are usually built to
intercross with existing main roads. For example, the existing roads
are often extended to new fields and branch roads are built from the
extension of these roads. This mechanism can lead to the emergence
of ``main roads" in urban system, and it is quite similar to the
``growth" and ``preferential attachment" in BA model.

\begin{figure}
\scalebox{0.8}[0.8]{\includegraphics{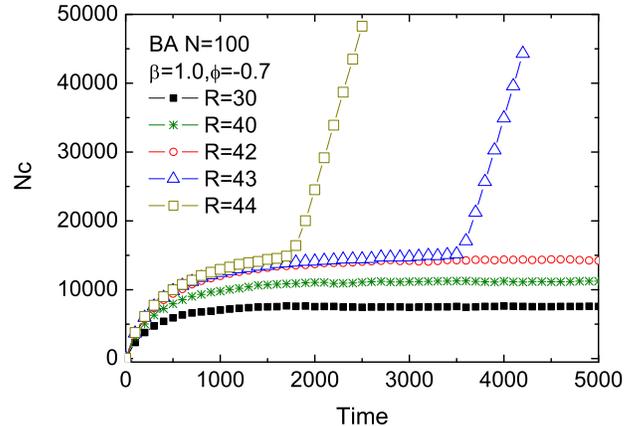}}
\caption{\label{BANc} (Color online)
Typical evolution of vehicle number in the scale-free system.
The critical generating rate of vehicles is $R_c=42$.}
\end{figure}

Figure \ref{BANc} shows the typical evolution of vehicle number in the system.
The same behavior as in the lattice case can be observed.
When $R<R_c$, the system is in free flow state, and the system will jam
when $R>R_c$.

\begin{figure}
\scalebox{0.8}[0.8]{\includegraphics{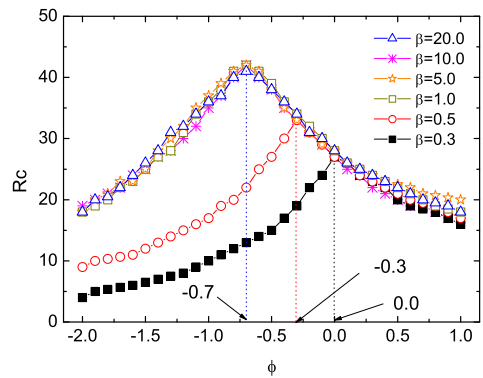}}
\caption{\label{BARc} (Color online)
The system's capacity vs $\phi$ with $V_{max}=100$.
The data are obtained by averaging $R_c$ over ten network realizations.}
\end{figure}

We first simulate the traffic on a network of $N=100$ nodes (roads)
with $m_0=m=5$.
This relatively small system can be seen as simulating the
backbone of a city¡¯s urban traffic network.
In such a scale-free network, it is very important to investigate
the effect of navigation strategy on the overall capacity.
Figure \ref{BARc} depicts the variation of $R_c$ with $\phi$.
It is shown that $R_c$ will be optimized at some typical value of
$\phi_c$. For the case of $V_{max}=100$, when $\beta$ is above one,
the system's overall capacity will be optimized when $\phi_c=-0.7$
and with the $R_c \approx 42$. When $\beta$ decreases below one, the
system's efficiency will decrease rapidly. And the optimal value of
$\phi$ will increase with the decrease of $\beta$. When $\beta=0.3$,
$\phi_c=0.0$, implies that the best strategy is
random-walk. 
We note that $\phi_c<0$ means to use the minor roads first.
Therefore, if intersection capacity and road condition are
the same for both main roads and minor roads, the best strategy
for the whole system is to encourage drivers to use minor roads
first.

\begin{figure}
\scalebox{0.8}[0.8]{\includegraphics{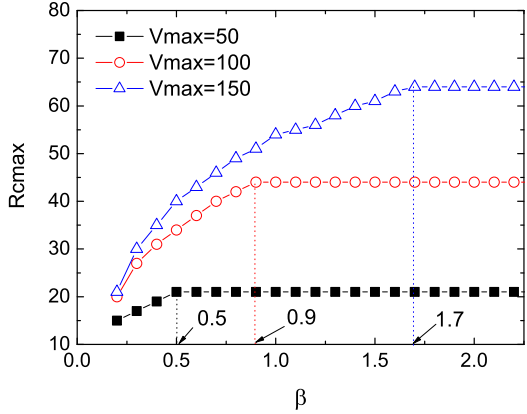}}
\caption{\label{BARcmB} (Color online)
The system's maximal capacity $R_{cmax}$ vs $\beta$
with different $V_{max}$.
}
\end{figure}

\begin{figure}
\scalebox{0.8}[0.8]{\includegraphics{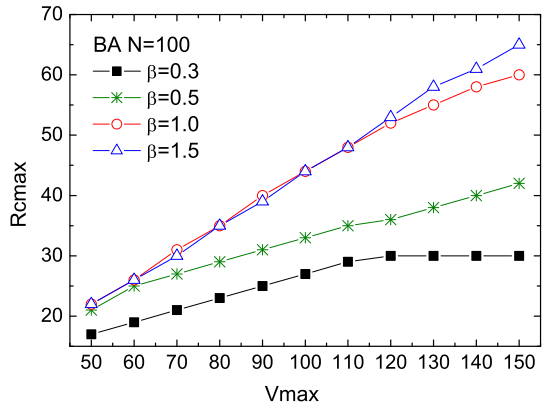}}
\caption{\label{BARcmV} (Color online)
The system's maximal capacity $R_{cmax}$ vs $V_{max}$
with different $\beta$.}
\end{figure}

Figure \ref{BARcmB} shows the variation of maximum value of $R_c$
(the peak value in Fig.\ref{BARc}) with the increment of $\beta$ for
different value of $V_{max}$. 
One can see that $R_{cmax}$ increase first and then come to saturation.
And in Fig.\ref{BARcmV}, the variation of $R_{cmax}$ with $V_{max}$
is shown.
The behaviors are similar with the lattice grid case.

\begin{figure}
\scalebox{0.8}[0.8]{\includegraphics{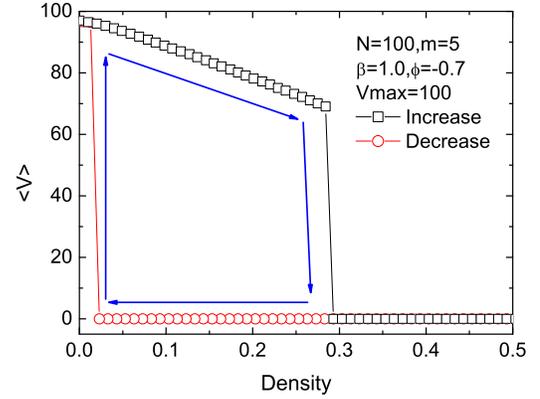}}
\caption{\label{BAVD} (Color online)
Average velocity vs density.
The black square line shows the velocity variation when
adding vehicles to the system (increase density),
while the red circle line shows the velocity variation
when drawing out vehicles from the system (decrease density).
The density values corresponding to the sudden transitions are
0.29 and 0.02 respectively.
The arrows are showing the hysteresis as a guide for the eyes.
}
\end{figure}

\begin{figure}
\scalebox{0.8}[0.8]{\includegraphics{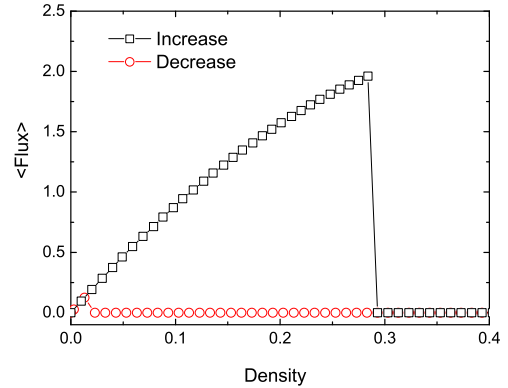}}
\caption{\label{BAFD} (Color online)
Flux vs density.
}
\end{figure}

\begin{figure}
\scalebox{0.8}[0.8]{\includegraphics{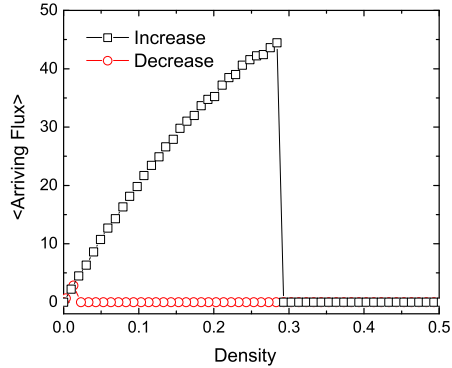}}
\caption{\label{BAFaD} (Color online)
Arriving flux vs density.
Arriving flux vs density in the cases of increasing and decreasing density.
Other parameters are the same as in Fig.\ref{BAVD}.}
\end{figure}

Finally, we try to reproduce the dependencies of average velocity
and traffic flux on vehicle density, which are important criteria
for evaluating the transit capacity of a traffic system. To simulate
the case of constant vehicle density, the number of arrived vehicles
at each time step is recorded, and the same number of vehicles are
added to the system at the beginning of next step. In
Fig.\ref{BAVD}, the velocity-density relation is displayed. Here the
vehicle density of the system is calculated with $\rho=N_c / C_t$.
The velocity firstly decreases gradually with the density. At the
density of $0.29$, the velocity suddenly drops to zero, indicating
the system enters the jam state. Two branches of the fundamental
diagram coexist between $0.04$ and $0.29$. The upper branch is
calculated by adding vehicles to the system (increasing density),
while the lower branch is calculated by randomly removing vehicles
from a jam state and allowing the system to relax after the
intervention (decrease density). In this way a hysteresis loop can
be traced (arrows in Fig.\ref{BAVD}). In the lower branch, the
velocity keep zero until the density is very low. This is because
some main roads are congested, thus the vehicles can not move on
these roads, and this state will not be alleviated by removing
vehicles randomly from the system.


Then we investigate the dependence of traffic flux on density.
Two kinds of traffic flux are studied: the movement flux and
the arriving flux.
The movement flux is calculated as the product of average
velocity and vehicle density.
It corresponds to the average number of vehicles passing a
given spot in the system per time step.
Figure \ref{BAFD} shows this flux-density relation of the system.
The arriving flux is calculated as the number of vehicles that
successfully reach their destination in each time step.
In Fig.\ref{BAFaD}, the relation of arriving flux vs traffic
density is shown.
In both cases, the hysteresis exist between the same values of
density as in Fig.\ref{BAVD}.
And the maximum arriving flux (45 vehicles per step) is
corresponding to $R_c=44$ when $V_{max}=100$ and $\beta=1.0$
as shown in Fig.\ref{BARcmB}.

The system's sudden drop to the jam state indicates
a first-order phase transition.
The phase transition and hysteresis can be explained
as follows.
According to the evolution rules, when a road is full of
vehicles, the velocity will be zero and the vehicles on
neighboring nodes can not turn to it.
So the vehicles may also accumulate on the neighboring
nodes and get congested.
This mechanism can trigger an avalanche across the system
when the density is high, thus a sudden phase transition
happen at this point.
As for the lower branch, starting from an initial congested
configuration, the system will have some congested roads
that are very difficult to dissipate.
These roads will decrease the system efficiency by
affecting the surrounding roads until all roads are
not congested, thus we get the lower branch.

\section{Conclusions and Discussions}

In conclusion, the urban traffic is simulated using a model
based on dual approach.
The model considers both the movement of vehicles on the road
and the navigation of drivers in the system.
The intersection and road conditions are naturally incorporated,
and their effects on the whole system's efficiency are investigated.
In a systemic view of overall efficiency, the model reproduces
several significant characteristics of urban traffic, such as
phase transition, hysteresis, velocity-density relation and
flux-density relation.
The simulation results are a kind of similar to that of the
BML model \cite{BML}, which was proposed in 1992 to simulate
the urban traffic on a square lattice.
But the present model is more general, and with more
interesting findings.

As comparing the simulation results on well-planned grid and on
self-organized scale-free network, one can see that the grid
network are more efficient than a self-organized one.
Given that the total capacity ($C_t \approx 57,600$ and $50,000$) is almost
the same for the two systems considered in this paper, the maximal number
of vehicles running on the grid is more than that on the scale-free
network (Fig.\ref{LNc} and \ref{BANc}), and the overall
capacity $R_c$ takes much larger value on the lattice grid
(Fig.\ref{RcB} and \ref{BARcmB}).
This is in agreement with the previous studies showing that
homogeneous networks can bear more traffic because of the
absence of high-betweenness nodes \cite{Gui,Tadic}.

When considering the navigation on a scale-free network,
the results are in agreement with previous studies that
the traffic system will be more efficient by avoiding the
central nodes \cite{Wang}.

The work can be extended further in many ways.
One can modify the model to capture more details in real traffic,
such as the role of traffic lights, the differences between main
roads and minor roads.
Better navigation strategies can also be coined in the dual perspective.
The resilience of traffic system against road failures
will be of great importance and research interest.

\section*{Acknowledgement}
This work is funded by National Basic Research Program of China
(No.2006CB705500), the NNSFC under Key Project Nos.10532060 and
10635040, Project Nos.70601026, 10672160, 10404025, the CAS
President Foundation, and by the China Postdoctoral Science
Foundation (No. 20060390179). Y.-H. Wu acknowledges the support of
Australian Research Council through a Discovery Project Grant.

\end{document}